\documentclass[
 reprint,
 amsmath,amssymb,
 nofootinbib,
 aps,
 prd,
 superscriptaddress,
 floatfix
]{revtex4-2}
\usepackage{graphicx}
\usepackage{dcolumn}
\usepackage{bm}
\usepackage{amsmath}
\usepackage{float}
\usepackage{lipsum}
\usepackage{xcolor}
\usepackage{times}
\usepackage{textcomp}
\usepackage{latexsym}
\usepackage[hidelinks]{hyperref}
\usepackage{acro}
\usepackage{mathtools}
\usepackage{url}
\usepackage[utf8]{inputenc}
\usepackage[normalem]{ulem}
\usepackage{xspace}
\usepackage{lineno}
\usepackage{scrextend}
\definecolor{lightgreen}{RGB}{179,255,164}
\definecolor{lightblue}{RGB}{196,206,255}
\definecolor{light-gray}{RGB}{230,230,230}
\definecolor{rb4}{HTML}{27408B}

\newcommand{\msun}{\ensuremath{M_{\odot}}\xspace}
\newcommand{\di}{\ensuremath{\mathrm{d}}\xspace}
\newcommand{\nLo}{\ensuremath{{n_{L,0}}}\xspace}

\newcommand{\pairing}{\ensuremath{\mathcal{P}}}
\newcommand{\HyperParamLens}{\ensuremath{\pmb{\Lambda}_L}}
\newcommand{\HyperParamSource}{\ensuremath{\pmb{\Lambda}_S}}
\newcommand{\HyperPriorFull}{\ensuremath{\pi_{\Lambda}\left(\HyperParamSource,\HyperParamLens\right)}}
\newcommand{\HyperPriorLens}{\ensuremath{\pi_{\Lambda}\left(\HyperParamLens\right)}}
\newcommand{\HyperPosFull}{\ensuremath{p_{\Lambda}\left(\HyperParamLens \vert \pmb{d}\right)}}

\newcommand{\GWlikelihood}{\ensuremath{L_{\rm gw}\left(d^i \vert \pmb{x}^i\right)}}
\newcommand{\GWposterior}{\ensuremath{p_{\rm{gw}}\left(\pmb{x}^i \vert d^i\right)}}
\newcommand{\DefaultGWPrior}{\ensuremath{\mathrm{Pr}\left(\pmb{x}^i\right)}}
\newcommand{\LensParam}{\ensuremath{\pmb{\Tilde{x}}_L}}
\newcommand{\SourceParam}{\ensuremath{\pmb{\Tilde{x}}_S}}
\newcommand{\LensParamFull}{\ensuremath{\pmb{x}_L}}
\newcommand{\PointMassLensParam}{\ensuremath{z_L,M_L}}

\newcommand{\SourceParamFull}{\ensuremath{\pmb{x}_S}}

\newcommand{\SourcePriorFullith}{\ensuremath{\pi_{S}\left(z_S^i, \SourceParam^i\right)}}
\newcommand{\LensPriorFull}{\ensuremath{\pi_{L}\left(y, \LensParamFull | z_S, \HyperParamLens, \pairing\right)}}
\newcommand{\LensPriorFullith}{\ensuremath{\pi_{L}\left(y^i, \LensParamFull^i | z_S^i, \HyperParamLens, \pairing\right)}}

\newcommand{\LensIntrinsicPriorFull}{\ensuremath{\pi_L^{\prime}\left(z_L, \LensParam| z_S, \HyperParamLens\right)}}

\newcommand{\yPriorFull}{\ensuremath{\pi_{y}\left( y | z_S, \HyperParamLens, \pairing\right)}}
\newcommand{\LensParamFullPriorFull}{\ensuremath{\pi_{\LensParamFull}\left(z_L, \LensParam | z_S, \HyperParamLens, \pairing\right)}}

\newcommand{\piSim}{\ensuremath{p_{\mathrm{sim}}\left(y|z_S, \HyperParamLens, \pairing \right)}}

\newcommand{\MLzPriorFull}{\ensuremath{\pi_{M_{Lz}}\left(M_{Lz} | z_S, \HyperParamLens, \pairing\right)}}
\newcommand{\MLPriorFull}{\ensuremath{\pi_{M_{L}}\left(M_{L} | z_S, \HyperParamLens, \pairing\right)}}
\newcommand{\zLPriorFull}{\ensuremath{\pi_{z_{L}}\left(z_{L} | z_S, \HyperParamLens, \pairing\right)}}

\newcommand{\MLIntrinsicPrior}{\ensuremath{\pi_{M_{L}}^{\prime}\left(M_{L}|\alpha_L\right)}}
\newcommand{\zLIntrinsicPrior}{\ensuremath{\pi_{z_{L}}^{\prime}\left(z_{L} | z_S\right)}}
\newcommand{\MLIntrinsicPriorFull}{\ensuremath{\pi_{M_{L}}^{\prime}\left(M_{L}|z_S, \HyperParamLens\right)}}
\newcommand{\zLIntrinsicPriorFull}{\ensuremath{\pi_{z_{L}}^{\prime}\left(z_{L} | z_S, \HyperParamLens\right)}}
\newcommand{\MLmin}{\ensuremath{M_{L,\rm min}}}
\newcommand{\MLmax}{\ensuremath{M_{L,\rm max}}}
\newcommand{\perMpcCube}{\ensuremath{\mathrm{Mpc}^{-3}}}

\begin{document}
\preprint{APS/123-QED}

\title{Inferring the Intermediate Mass Black Hole Number Density from Gravitational Wave Lensing Statistics}
\author{Joseph Gais}
\email{1155138494@link.cuhk.edu.hk}
\affiliation{%
Department of Physics, The Chinese University of Hong Kong, Shatin, N.T., Hong Kong.
}%
\author{Ken Ng}
\affiliation{LIGO, Massachusetts Institute of Technology, Cambridge, Massachusetts 02139, USA}
\affiliation{Kavli Institute for Astrophysics and Space Research, Department of Physics, Massachusetts Institute of Technology, Cambridge, Massachusetts 02139, USA}
\author{Eungwang Seo}
\affiliation{%
Department of Physics, The Chinese University of Hong Kong, Shatin, N.T., Hong Kong.
}%

\author{Kaze W.K. Wong}
\affiliation{%
Center for Computational Astrophysics, Flatiron Institute, New York, NY 10010, USA
}

\author{Tjonnie G. F. Li}
\affiliation{%
Department of Physics, The Chinese University of Hong Kong, Shatin, N.T., Hong Kong.
}
\affiliation{Institute for Theoretical Physics, KU Leuven, Celestĳnenlaan 200D, B-3001 Leuven, Belgium.}
\affiliation{Department of Electrical Engineering (ESAT), KU Leuven, Kasteelpark Arenberg 10, B-3001 Leuven, Belgium}
\begin{abstract}
The population properties of intermediate mass black holes remain largely unknown, and understanding their distribution could provide a missing link in the formation of supermassive black holes and galaxies.
Gravitational wave observations can help fill in the gap from stellar mass black holes to supermassive black holes.
In our work, we propose a new method for probing lens populations through lensing statistics of gravitational waves, here focusing on inferring the number density of intermediate mass black holes.
Using hierarchical Bayesian inference of injected lensed gravitational waves, we find that existing gravitational wave observatories at design sensitivity could either identify an injected number density of $10^6  \mathrm{Mpc}^{-3}$ or place an upper bound of $\lesssim 10^4 \mathrm{Mpc}^{-3}$ for an injected $10^3   \mathrm{Mpc}^{-3}$.
More broadly, our method could be applied to probe other forms of compact matter as well.

\end{abstract}
\maketitle
\section{Introduction}
To date, we have detected dozens of black holes within the stellar mass range $\mathcal{O}(1-100)M_\odot$ from binary black hole merger gravitational wave emission \cite{abbott2019gwtc,abbott2020gwtc, abbott2021gwtc} and X-ray binary observations \cite{mcclintock2003black, remillard2006x}, as well as supermassive black holes of mass $> \mathcal{O}(10^6) M_\odot$, first identified from stellar orbits about the center of the Milky Way \cite{ghez2005stellar} and now imaged by the Event Horizon Telescope \cite{event2019first, akiyama2019first2, akiyama2019first3, akiyama2019first4, akiyama2019first5, akiyama2019first6}.
The least understood parameter space of black holes lies between these two ranges, the so-called intermediate mass black holes (IMBH) in the mass range $[10^2, 10^6] M_\odot$.
Understanding the formation channels of supermassive black holes and galaxies themselves will require filling in the missing link of IMBHs.

IMBHs may soon be detected.
Search methods include stellar and gas dynamical searches as well as accreting IMBHs within galactic nuclei suggest a number of tentative IMBH discoveries (see \cite{greene2019intermediate} for a recent review). 
Recently, the first half of LIGO-Virgo’s third observing run has detected the gravitational waves of a binary black hole merger with a remnant mass of $142 M_\odot$ \cite{abbott2020gw190521}, the first ever confirmed IMBH.
In addition to measurements of IMBH remnants, another possible method for detecting IMBHs lies in measuring gravitational wave lensing effects.

If a gravitational wave passes by an IMBH mass lens closely, the measured gravitational wave will have a frequency dependent amplification factor altering the waveform \cite{takahashi2003wave}. 
From careful study of detected gravitational waves, we may determine the lens parameters, with recent work demonstrating the detection of mass of an IMBH lens \cite{lai2018discovering} and how gravitational wave lensing can constrain black hole populations \cite{diego2020constraining}.
Although no gravitational wave event has yet been conclusively identified as being lensed \cite{hannuksela2019search, liu2021identifying, mcisaac2020search, abbott2021search}, tentative lensing rates estimates suggest aLIGO could detect $\mathcal{O}(1) \mathrm{yr}^{-1}$ lensed events at design sensitivity \cite{oguri2018effect,ng2018precise,li2018gravitational}.

Building off of \cite{lai2018discovering}, we consider the lensing of gravitational waves by IMBHs as a means of inferring the IMBH number density $n_L$.
We develop an analytical model verified by simulation results for the distribution of the single-lensing event parameters, the normalized impact parameter $y$ and redshifted lens mass $M_{lz}$.
We then use a hierarchical Bayesian model for constraining possible $n_L$ values from a population of recovered $y$'s alongside our simulated distributions of impact parameter for different lens number density.
Since a priori we have no means of identifying a lensed gravitational wave, we conduct the parameter estimation on all gravitational wave events, where the posterior of unlensed gravitational waves should demonstrate significant support at large $y$ and little support at $y \lesssim \mathcal{O}(1)$.
In contrast, lensed gravitational waves with $y \lesssim 1$ should be recovered from the parameter estimation.
For any gravitational wave event, we conduct parameter estimation of the redshifted lens mass, $M_{lz}$ and $y$. 
The set of lens parameter estimation allows us to build a distribution for the full population of $y$ values.
In turn, we are able either to constrain the number density of IMBHs if no IMBH mass range lenses are present within the full population, or measure on the IMBH number density if IMBH lens events are detected.

Injecting a catalog of $\sim 200$ events drawn from $n_L = \{10^3, 10^6\} \mathrm{Mpc}^{-3}$ with a design sensitivity LIGO Hanford, LIGO Livingston \cite{aasi2015advanced} and Virgo \cite{acernese2014advanced} observatory network, we can confidently detect the density of IMBH lenses at $10^6 \mathrm{Mpc}^{-3}$ or constrain to $\lesssim 10^4 \mathrm{Mpc}^{-3}$ for a number density of $10^3 \mathrm{Mpc}^{-3}$, on the scale of IMBH densities inferred from gamma ray burst observations \cite{paynter2021evidence}.
Combining measurements from lensing statistics as well as with parameter estimation of source masses in gravitational wave mergers could then shed light on the largely unknown population of IMBH lenses.

We begin by describing the effect of a point mass lens on a gravitational wave in Sec. \ref{sec:gwlensing}. Then, in Sec. \ref{sec:hba}, we derive a hierarchical Bayesian model to infer the point mass lens population from detected gravitational wave events. In Sec. \ref{sec:nearest_effective}, we detail an analytical population model for IMBH lenses, validating our model against simulated results. We then conduct an injection campaign in the LIGO-Virgo detector network as described in Sec. \ref{sec:gwlens_pe}. Finally, in Sec. \ref{sec:results}, we present the recovered lens number density from our injections, and discuss our results and impact of improved detector networks on probing the IMBH population in Sec. \ref{sec:discussion}.

\section{Gravitational Wave Lensing}~\label{sec:gwlensing}

When a gravitational wave passes by a massive object, it is lensed in a manner similarly to electromagnetic waves.
In the geometric optics regime, i.e., when the dimensionless frequency $w = 8 \pi M_{Lz} f \gg 1$, where $M_{Lz}$ is the redshifted lens mass with gravitational frequency $f$ in the detector's frame, the amplitude of the gravitational wave is either magnified or demagnified while the phase content remains unchanged.
However, in the wave optics regime where $w \lesssim 1$, both the amplitude and phase of the gravitational wave are modulated in a frequency-dependent manner, yielding a rich structure in the lensed gravitational wave.
Lensed gravitational waves could soon be detected \cite{hannuksela2019search, ng2018precise, oguri2018effect}, with applications ranging from improved sky localization \cite{hannuksela2020localizing}, tests of the polarization of gravitational waves \cite{goyal2021testing}, or probing dark matter \cite{urrutia2021lensing}.

\begin{figure}
    \centering
    \includegraphics[width=0.5\textwidth]{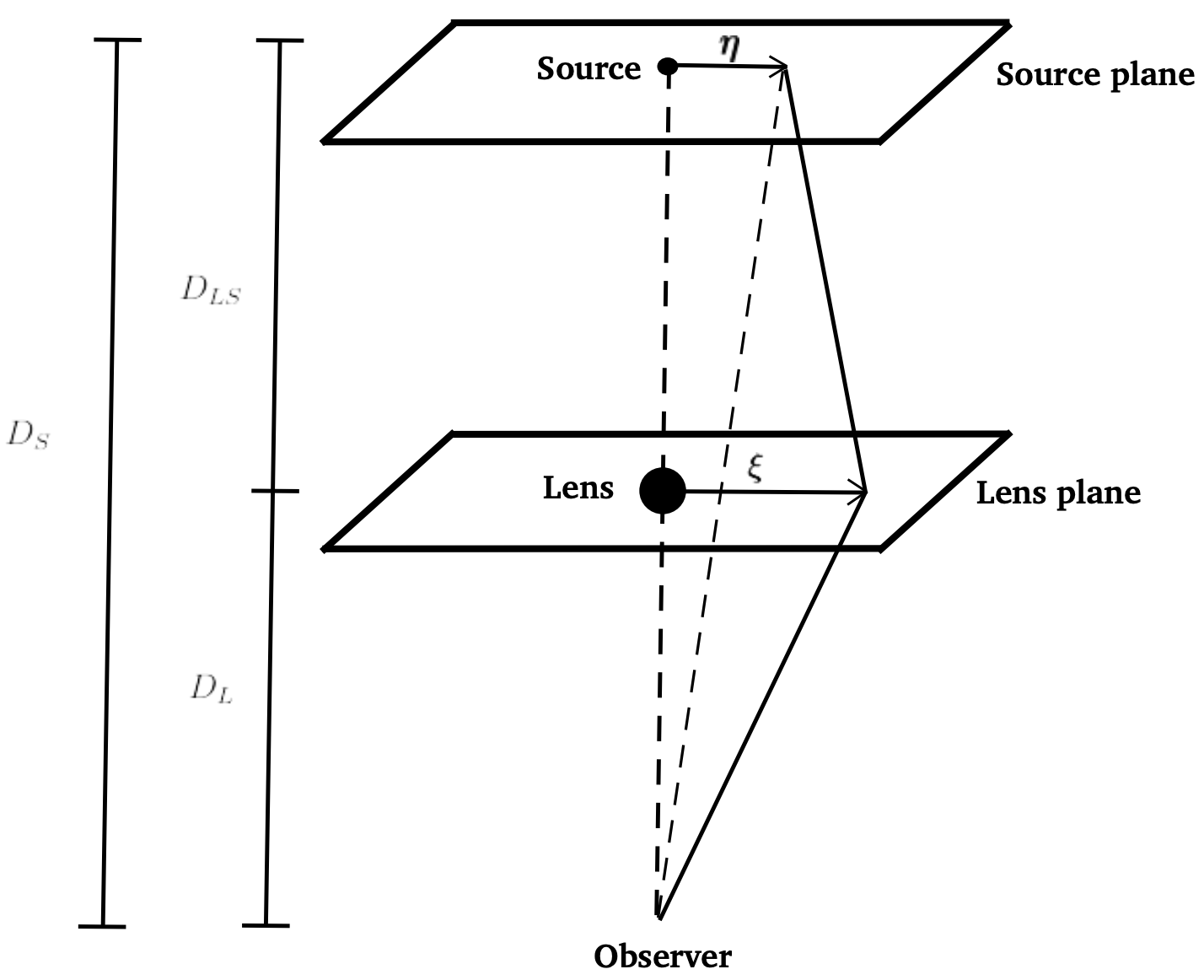}
    \caption{Basic lensing geometry for a gravitational wave lensed by a point mass in the thin lens approximation. In the plane of the sky with the lens at the origin, the source is located at $\pmb{\eta}$, passes the lens plane with impact parameter $\pmb{\xi}$, and then deflected by the lens at the lens plane, ultimately reaching the observer. $D_S$ signifies the angular diameter distance from the observer to the source, $D_L$ is the angular diameter distance from observer to lens, and $D_{LS}$ is the angular diameter distance from lens to source, which is \textit{not} equal to $D_S - D_L$.}
    \label{fig:lens_schematic}
\end{figure}

Here, we focus on the case of a gravitational wave lensed by a single point mass, illustrated in Fig. \ref{fig:lens_schematic}.
The details of the analytical calculation for the lensing amplification factor are outlined in App.~\ref{appendix:a}, resulting in an analytical solution for the isolated point mass,
\begin{align}
\label{eq:AFpointmass}
F(w)&=\exp\bigg{\{}\dfrac{\pi w}{4}+i \dfrac{w}{2} \bigg{[}\text{ln} \bigg{(}\dfrac{w}{2}\bigg{)}-2\phi_m(y)\bigg{]}\bigg{\}} \nonumber \\
&\times \Gamma \bigg{(}1- \dfrac{i}{2}w\bigg{)} {_1}F_1\bigg{(}\dfrac{i}{2}w, 1; \dfrac{i}{2}w y^2\bigg{)},
\end{align}
where $w=8 \pi M_{Lz}f$ is the dimensionless frequency, $y$ is the impact parameter normalized by the lens' Einstein radius, $M_{Lz}$ is the redshifted lens mass, $_1F_1$ is the confluent hypergeometric function, and
\begin{align}
&\phi_m(y)=\frac{(x_m-y)^2}{2} - \ln x_m, \\
&x_m=\frac{y+\sqrt{y^2+4}}{2}.
\end{align}
The lensed waveform is then,
\begin{equation}
    \psi^L(f) = F(f) \psi_0(f)
\end{equation}
where $\psi_0(f)$ is the frequency-domain base waveform and $F(f)$ is the amplification factor.

Previous studies demonstrate that the gravitational wave event parameters and lens parameters, $M_{Lz}$ and $y$, the impact parameter of the source-lens pair normalized by the lens' Einstein radius, are detectable from Bayesian parameter estimation of the lensed gravitational wave for IMBHs \cite{lai2018discovering}.
Following this example, we prepare a likelihood model for a lensed gravitational wave, from which one can infer the posterior on the lensing parameters.
When $w y^2 /2 \gg 1$, the amplification factor is highly oscillatory in the frequency domain, the geometric optics approximation can be used.
Using a dynamical lookup table in $(w, wy^2 /2)$ for the evaluation of the hypergeometric function in $F(f)$, we are able to rapidly evaluate the amplification factor such that lensing parameter estimation is feasible, and use the geometric optics approximation for $w y^2 > 1000$ elsewhere.

\section{Hierarchical Bayesian Analysis}\label{sec:hba}
In this section, we list the mathematical details of the hierarchical inference model for a generic lensing scenario.
We seek to measure the properties of the lens population parameterized by $\HyperParamLens$.
Given a dataset $\pmb{d}=\{d^i\}$ of $N$ detections and the properties of source population parameterized by $\pmb{\Lambda}_S$, we can compute the posterior of $\HyperParamLens$, $p_{\Lambda}(\HyperParamLens \vert \pmb{d}, \pmb{\Lambda}_{S})$,  by combining the measurement of waveform parameters $\pmb{x}$ of each detection,
\begin{equation}
    \label{eq:hba_full}
    \frac{p_{\Lambda}(\HyperParamLens \vert \pmb{d}, \pmb{\Lambda}_{S})}{\HyperPriorFull} \propto \prod_{i=1}^N\int \GWlikelihood \pi_{\rm gw}(\pmb{x}^i \vert \pmb{\Lambda}_S, \HyperParamLens)\mathrm{d}\pmb{x}^i,
\end{equation}
where $\GWlikelihood$ is the likelihood of the $i$-th gravitational wave detection, $\pi_{\rm gw}(\pmb{x}^i \vert \pmb{\Lambda}_S, \HyperParamLens)$ is the distribution of waveform parameters given both the source and lens population properties, and $\HyperPriorFull$ is the prior of $(\HyperParamSource,\HyperParamLens)$.
While one can simultaneously infer $(\HyperParamSource,\HyperParamLens)$, we expect that the population properties of sources and lenses are weakly correlated and leave out $\HyperParamSource$ for the rest of the paper for simplicity.
We list our choice of source population properties, such as BBH mass spectrum and redshift evolution, in App.~\ref{app:source_pop}.
In the following, we separate the waveform parameters into $(y,\SourceParamFull, \LensParamFull)$, in which $y$ can be thought of the parameter characterizing the pairing of a source and a lens, $\SourceParamFull=(z_S, \SourceParam)$ is the set of source parameters including source redshift $z_S$ and other parameters irrelevant to lensing, $\SourceParam$, and $\LensParamFull=(z_L, \LensParam)$ is the set of lensing-relevant parameters including the lens redshift $z_L$ and the model-dependent parameters characterizing the internal properties of the lens, $\LensParam$.
nt parameters given hyperparameters $\pmb{\Lambda}$ which we simulate directly.

We expect that $\SourceParam$ and $\LensParamFull$ are independent of each other and hence their distributions are separable.
We treat the constraint that a lens must be inside the volume within $z_S$, $z_L<z_S$ as a condition imposed on the lens distribution in Bayes' theorem.
One can further marginalize over other irrelevant source parameters $\SourceParam$.
Putting these steps together, Eq.~\eqref{eq:hba_full} becomes
\begin{align}
\label{eq:separated_likelihood}
    \frac{\HyperPosFull}{\HyperPriorLens}  \propto \prod_{i=1}^N &\iiiint \big[ \GWlikelihood \LensPriorFullith \nonumber \\
    &\times \SourcePriorFullith \di z_S^i \di y^i \di \LensParamFull^i  \big] \di \pmb{\Tilde{x}}_S^i,
\end{align}
where $\LensPriorFull$ is the distribution of lens parameters given a source at redshift $z_S$, and $\pi_S$ is the prior of the source parameters.
The conditional statement $\pairing$ denotes the requirement of a source-lens pair having the strongest diffraction along the line of sight.
We will explain the importance of this notion in Sec.~\ref{sec:nearest_effective}.

To evaluate Eq.~\eqref{eq:separated_likelihood}, we can use importance sampling by recognizing that $\GWlikelihood \DefaultGWPrior = \GWposterior$, where $\DefaultGWPrior$ is the prior of waveform parameters used in the parameter-estimation algorithm that estimates the posterior of waveform parameters, $\GWposterior$.
We can reweigh the samples drawn from the estimated posterior to evaluate the hierarchical likelihood,
\begin{align}
\label{eq:summed_likelihood}
    \frac{\HyperPosFull}{\HyperPriorLens}  \propto \prod_{i=1}^N \bigg\{ \frac{1}{K^i} & \sum_{j=1}^{K^i} \bigg[ \frac{\pi_{S}(z_S^{i,j}, \SourceParam^{i,j})}{\mathrm{Pr}(y^{i,j},z_S^{i,j},\SourceParam^{i,j},\LensParamFull^{i,j})}  \nonumber \\
    & \times  \pi_{L}(y^{i,j},\LensParamFull^{i,j}\vert z_S^{i,j}, \HyperParamLens, \pairing) \bigg]\bigg\},
\end{align}
where $(\cdot)^{i,j}$ denote the $j$-th sample drawn from $K^i$ posterior samples of the $i$-th event.

Generically, in hierarchical Bayesian analysis of hyperparameters, the selection bias must be taken into account.
For $y \ll 1$, the lensed waveform is greatly amplified~\cite{nakamura1998gravitational,nakamura1999wave, takahashi2003wave}, resulting in higher SNR values.
Selection of only those events above a certain threshold will then bias the recovered hyperparameter posterior towards higher lens number densities, as events with higher $y$ values (and thus, less of a lensing effect) are less likely to have a sufficiently high SNR.
However, for the physically motivated regime of number densities we consider, $y \gg 1$ in most events, resulting in magnifications very close to unity, and so the SNR of any event is hardly affected by lensing (and by extension the lens number density).
Thus, the SNR selection is unlikely to bias our results and we ignore it for simplicity.

\section{Distribution of the nearest-effective lenses}\label{sec:lenspopulation}
\subsection{Notion of the nearest-effective lens}\label{sec:nearest_effective}

We observe the population of the source-lens systems rather than the population of isolated lenses.
One needs to cautiously account for this subtle difference when modeling $\pi_L$ in Eq.~(\ref{eq:summed_likelihood}), which is no longer the intrinsic distribution of the lenses.
We assume that a source is solely diffracted by a single lens, i.e., multiple lensing due to the next neighboring lenses is negligible.
Since the size of the Einstein ring also affects the magnitude of $y$, the nearest-neighbor lens (i.e. with the smallest value of $\theta_S = \eta/D_S$) does not necessarily give rise to the strongest effect of diffraction.
Instead, a source is the most diffracted by a lens whose parameters result in the smallest value of $y$.
We call such lenses as the \textit{nearest-effective} lenses.
In terms of the lensing statistics, the statement $\pairing$ is equivalent to the requirement of minimum $y$ when pairing the lenses and sources.
We can model the nearest-effective pairing by characterizing the distribution of neighboring lenses through a spatial Poisson process, which only depends on the spatial distribution among the lenses but not on the internal properties of the lenses.
This is achievable by considering $y$ as an \textit{effective distance} between a source and its nearest effective lens on the sky plane.
Assuming the lenses are uniformly distributed on the sky plane, we can separate the joint distribution of $y$ and $\LensParamFull$ into
\begin{align}\label{eq:pi_y_Mlz}
    &\quad \LensPriorFull \nonumber \\
    &= \yPriorFull \LensParamFullPriorFull,
\end{align}
where $\pi_{y}$ and $\pi_{\LensParamFull}$ are the distributions of $y$ and $\LensParamFull$ conditioned on the nearest-effective pairing between sources and lenses, respectively.
In the following, we first derive $\pi_y$ and $\pi_{\LensParamFull}$ from the spatial Poisson process, then list out the mathematical details in the case of point-mass lenses, and validate the analytical model by comparing it to the direct simulation of the nearest-effective pairing of the source-lens systems.

\subsection{Spatial Poisson Process}\label{sec:spatialPoisson}
With a source centered at the origin, the probability that there are $k$ lenses within an effective distance $y$ is
\begin{equation}
    \text{Poisson}(k|\Sigma) = \frac{\left(\Sigma \pi y^2\right)^k}{k!}e^{-\Sigma \pi y^2},
\end{equation}
where $\Sigma \pi $ is the effective density parameter of lenses within the volume of $z_S$ projected on the sky.
The differential probability of finding the nearest-effective lens inside an infinitesimal ring between $y$ and $y+\di y$ is the product of the probability that there is no lens within the circle of radius $y$, $\text{Poisson}(0|\Sigma) = e^{-\Sigma \pi y^2}$, and the probability of a lens lying inside the ring, $2\Sigma \pi y  \di y$.
Dividing this probability by $\di y$, the probability density function of the nearest-effective lens locating at $y$ is
\begin{equation}\label{eq:spatial_poisson}
    p(y) = 2 \Sigma \pi y  e^{-\Sigma \pi y^2}.
\end{equation}

Since $y$ is the dimensionless ratio of the angular separation between the source and the lens to the angular size of the lens Einstein ring, the effective density parameter can be interpreted as the mean fractional area of all lenses within $z_S$ relative to the full sky plane (or, equivalently, the inverse of the mean of $y^2$), i.e.,
\begin{align}~\label{eq:Sigma_def}
    \Sigma(z_S,\HyperParamLens) \pi = N_L(z_S)\frac{\pi \left\langle\theta_E^2 \right\rangle_{\HyperParamLens}}{4\pi},
\end{align}
where $N_L(z_S)=\int_0^{z_S} n_L(z_L) \di V_c(z_L)$ is the total number of lenses within the comoving volume $V_c(z_S)$ for an arbitrary number density evolution of lenses $n_L(z_L)$, and 
\begin{align}\label{eq:mean_thetaE}
    \left\langle\theta_E^2 \right\rangle_{\HyperParamLens} = \int \theta^2_E\left(z_L, \LensParam | z_S\right) \LensIntrinsicPriorFull \di z_L \di \LensParam
\end{align}
is the mean area enclosed by the Einstein rings, with $\langle \cdot \rangle_{\HyperParamLens}$ being the mean quantity over the intrinsic lens distribution parameterized by ${\HyperParamLens}$, $\LensIntrinsicPriorFull$ is the joint distribution of redshift and mass of the intrinsic lens population (i.e. regardless of the pairing with the sources).
Thus, the term $\Sigma \pi y^2$ in the exponent of Eq.~\eqref{eq:spatial_poisson} is equivalent to the mean number of lenses within the area $\pi \theta_S^2$.
The desired $\pi_y$ is then
\begin{align}\label{eq:pi_y}
    \yPriorFull = 2 \pi y \Sigma(z_S, \HyperParamLens) e^{-\Sigma(z_S, \HyperParamLens)\pi y^2}.
\end{align}

The pairing requirement, $\pairing$, favors a source-lens system with the largest $\theta_E$ to minimize the value of $y$.
One can think of the pairing condition as choosing the lens with the largest area, $\pi \theta_E^2$.
As a result, the final distribution of lens parameters in the source-lens systems has an additional \textit{lensing bias factor} proportional to $\theta_E^2$ for sources at the same $z_S$.
Mathematically, the distribution of $\LensParamFull$ after the nearest-effective pairing is
\begin{align}
\label{eq:pi_ML_zL}
    \LensParamFullPriorFull \propto \theta_E^2 \LensIntrinsicPriorFull,
\end{align}
which is indeed the integrand of Eq.~\eqref{eq:mean_thetaE}.

\subsection{Lensing Statistics for Point-mass Lenses}\label{sec:PointMassLens}

IMBHs with masses of $\sim \mathcal{O}(100-10^4)~\msun$ may serve as point mass lenses to diffract gravitational waves.
The mass profile of a point mass lens is entirely parameterized by its mass $M_L$, i.e., $\LensParam=M_L$.
Throughout the study, we assume the intrinsic lens mass spectrum does not evolve with lens redshift, i.e., $\pi_L^{\prime}=\pi_{M_L}^{\prime}\pi_{z_L}^{\prime}$, where $\pi_{M_L}^{\prime}$ and $\pi_{z_L}^{\prime}$ are the one-dimensional intrinsic distribution of lens mass and lens redshift, respectively.
We use a power-law mass spectrum with an index $\alpha_L$, $\MLIntrinsicPrior \propto M_{L}^{-\alpha_L}$, in the domain $[\MLmin=100~M_\odot, \MLmax=20000~M_\odot]$.
For simplicity, we keep the lens number density constant in the comoving frame such that the prior of lens redshift is $\zLIntrinsicPrior \propto \di V_c(z_L)/\di z_L$ for $z_L<z_S$.
We note that one can relax the assumption of constant density to infer the lens redshift evolution.
As such, we only have two hyperparameters, $\HyperParamLens=(\nLo, \alpha_L)$.

Now, we write down the expressions for $\pi_{\LensParamFull} \equiv \pi_{z_L} \pi_{M_L} $ and $\Sigma$.
Including the lensing bias factor, $\theta_E^2 \propto M_L D_{LS}/D_L$ at a fixed $z_S$, we have
\begin{align}
    & \zLPriorFull \propto \zLIntrinsicPriorFull \frac{D_{LS}}{D_L}, \\
    & \MLPriorFull \propto \MLIntrinsicPriorFull M_L.
\end{align}
Since $F(f)$ only depends on $(y, M_{Lz})$ and $z_L$ is not directly measured, we further marginalize $\pi_{M_L}\pi_{z_L}$ over $z_L$ to obtain the distribution of redshifted lens mass,
\begin{align}\label{eq:pi_MLz}
    &\quad \MLzPriorFull  \nonumber \\
    &\propto \int_0^{z_S} \left(\frac{M_{Lz}}{1+z_L}\right)^{1- \alpha_L} \frac{D_{LS}}{D_L} \frac{\di V_c}{\di z_L} \frac{\di z_L}{1+z_L},
\end{align}
for $M_L\in [\MLmin,\MLmax]$, and is zero otherwise.
The extra factor of $(1+z_L)^{-1}$ comes from the transformation of the differential $\di M_{Lz}=(1+z_L)\di M_L$.
Finally, the expression of $\Sigma$ for $\pi_y$ is
\begin{align}\label{eq:Sigma_zS}
    &\quad \Sigma(z_S, \HyperParamLens) = \frac{4\nLo\chi_S^3}{3 D_S} \left\langle M_L \right\rangle_{\HyperParamLens} \left\langle \frac{D_{LS}}{ D_L} \right\rangle_{\HyperParamLens},
\end{align}
where $\chi_S$ is the comoving distance at $z_S$, $\left\langle M_L \right\rangle_{\HyperParamLens}$ is the mean lens mass,
\begin{align}
    \left\langle M_L \right\rangle_{\HyperParamLens} = 
    \begin{cases}
        \dfrac{\MLmax-\MLmin}{\ln{\left(\MLmax/\MLmin\right)}} &\text{for~$\alpha_L= 1$} \\
        \\
        \dfrac{\ln{\left(\MLmax/\MLmin\right)}}{\MLmin^{-1}-\MLmax^{-1}} &\text{for~$\alpha_L= 2$} \\
        \\
        \dfrac{1-\alpha_L}{2-\alpha_L}\dfrac{\MLmax^{2-\alpha_L}-\MLmin^{2-\alpha_L}}{\MLmax^{1-\alpha_L}-\MLmin^{1-\alpha_L}} &\text{otherwise,}
    \end{cases}
\end{align}
and $\left\langle D_{LS}/ D_L \right\rangle_{\HyperParamLens}$ is the mean distance factor given by
\begin{align}
    \left\langle \frac{D_{LS}}{ D_L} \right\rangle_{\HyperParamLens} = \int_0^{z_S} \frac{D_{LS}}{ D_L} \frac{\di V_c}{\di z_L} \di z_L.
\end{align}
We use Planck 18 cosmology~\cite{aghanim2020planck} for the evaluation of cosmological distances.

\subsection{Validation}
Let us examine the behavior of $\pi_L$.
First, the inverse of the density parameter $\left(\Sigma\pi\right)^{-1}$ characterizes the scale of $y$.
In particular, the most probable value of $y$ (or the peak of $\pi_y$) is $y_p = \left(2 \Sigma \pi\right)^{-1/2}$.
This can be understood physically by interpreting $\left(\Sigma\pi\right)^{-1}$ as the ratio of the mean cross-section area, $\pi \left\langle \theta^2_S \right\rangle_{\HyperParamLens} \equiv 4\pi/N_L$, to the mean area of lenses, $\pi \left\langle \theta^2_E \right\rangle_{\HyperParamLens}$ (cf Eq.~\eqref{eq:Sigma_def}).
Second, in the limit of $y\to \infty$, the Gaussian term $e^{-\Sigma \pi y^2}$ regulates the linear increase in $\pi_L$ with $ye^{-\Sigma \pi y^2}\to 0$.
The impact parameter cannot be arbitrarily large because the separation between adjacent lenses is characterized by the scale of $\left( \Sigma \pi\right)^{-1/2}$.
Third, we consider the limit of $0<y<y_{\rm max}$, where $y_{\rm max}$ is the cut-off of $y$ satisfying $y_{\rm max}\ll (\Sigma\pi)^{-1/2}$.
In such limit, sources are distributed uniformly around the vicinity of the nearest-effective lens, resulting in a linear distribution of $y$.
Indeed, the spatial Poisson piece, $2\pi y\Sigma e^{-\Sigma \pi y^2}$, is well approximated by $2y/y_{\rm max}^2$ for $y^2\ll \left(\Sigma\pi\right)^{-1}$ and independent of $\Sigma$.
Together with the lensing bias factor $\propto \theta_E^2$, the asymptotic form of $\pi_L$ for $y\ll \left(\Sigma\pi \right)^{-1/2}$ is
\begin{align}
    &\quad \pi_L^{0}\left(y, z_L, \LensParam | z_S, \HyperParamLens, \pairing, y\ll \left(\Sigma\pi \right)^{-1/2} \right ) \nonumber \\ 
    & \propto \frac{2y}{y_{\rm max}^2} \theta_E^2 \LensIntrinsicPriorFull,
\end{align}
which, after the marginalization over $z_L$, recovers the usual definition of the \textit{lensing optical depth} (or the lensing probability) defined in the existing literature~\cite{turner1984statistics} for non-evolving point-mass lens distribution,
\begin{align}\label{eq:opticaldepth}
    \frac{\di^2 \tau}{\di y \di M_{L}} = \int_0^{z_S} 2  y \pi \theta_E^2 n_L(z_L) \frac{\di V_c}{\di z_L} \pi_{M_L}^{\prime}(M_L| z_S , \HyperParamLens) \di z_L,
\end{align}
up to some overall constants as $n_L(z_L) \di V_c/\di z_L \propto \pi_{z_L}^{\prime} $ and $\pi_L^{0}$ is a normalized probability density function rather than a probability function for the optical depth.

To test that the spatial-Poisson process accurately models the lensing statistics described thus far, we directly simulate a population of lenses and sources for a fixed value of $\nLo = 1000~\mathrm{Mpc}^{-3}$.
Lenses are placed uniformly in the plane, with a redshift distribution uniform in comoving volume, and have a power-law mass distribution with $\alpha_L = 1$ between $M_{L,\rm min}=100~\msun$ and $M_{L,\rm max}=20000~\msun$.
Source redshifts are assumed to follow the Madau-Dickinson star formation rate.
We then compute the $y$ value for each possible lens-mass pair, subject to the constraint that $z_S > z_L$.

\begin{figure}
    \centering
    \includegraphics[width=0.5\textwidth]{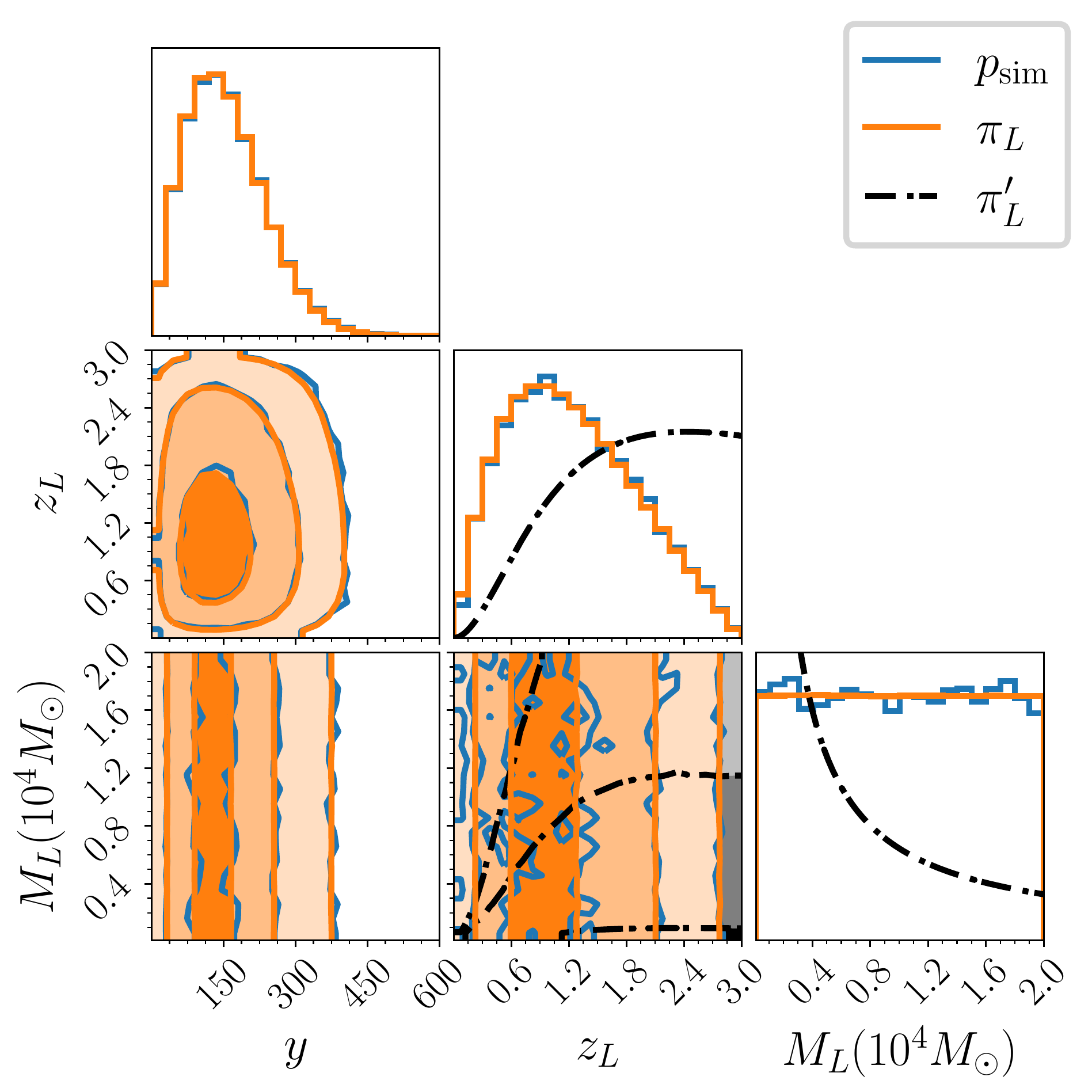}
    \caption{\label{fig:corner_sim}Corner plot of $(y,\PointMassLensParam)$ distributions simulated directly (blue), without bias factor (dashed black line), and with bias factor (orange), for $\alpha_L=1$.
    Because of the lensing bias from selecting source-lens pairs with the smallest $y$ value, the selected lens mass and lens redshift distributions, $\zLPriorFull \MLPriorFull$, are different from their pre-selection distribution, $\zLIntrinsicPriorFull \MLIntrinsicPriorFull$, and direct simulations confirm our bias factor.
    After selection, the lens redshift distribution now scales as $\frac{D_{LS}}{D_L} \frac{\di V_c}{\di z_L}$ and $\MLPriorFull$ is uniform.
    Furthermore, independent sampling of $(y,\PointMassLensParam)$ (orange contours) align with the direct simulation contours (blue), and so the 1D distributions are indeed uncorrelated following selection of minimum $y$.}
\end{figure}

We can identify that our bias factor described in Sec.~\ref{sec:spatialPoisson} is correct with the aid of a corner plot of our simulation in $(y, \PointMassLensParam)$.
Figure~\ref{fig:corner_sim} shows the corner plot with a fixed source redshift of $z_S = 3$ after selecting source-lens pairs, with the simulated marginalized distributions (blue), bias-factored analytical model (orange), and model without bias factoring (dashed black lines).
The spatial Poisson distribution $\pi_L(y)$ matches the simulated distribution $\piSim$ closely, validating the analytical model.
We note that the distributions $\MLPriorFull$ and $\zLPriorFull$ are altered from their pre-selection distribution, $\MLIntrinsicPriorFull$ and $\zLIntrinsicPriorFull$, with the bias-factored distributions matching the simulated distributions.
After selecting, $\pi_{M_L}$ is now uniform, and so $\pi_{M_{Lz}}$ follows the approximate shape of $\pi_{z_L}$.
The lens redshift distribution $\pi_{z_L}$ is more skewed towards smaller redshifts, as the bias factor $D_{LS}/D_L$ is maximized at smaller lens redshifts.
Additionally, drawing independent samples from the bias-factored distributions, plotted in orange contours, we find that they match the simulated contours, indicating that the lensing parameters $(y, \PointMassLensParam)$ are independent following selection of nearest effective lens-source pairs.

\begin{figure}
    \centering
    \includegraphics[width=\columnwidth]{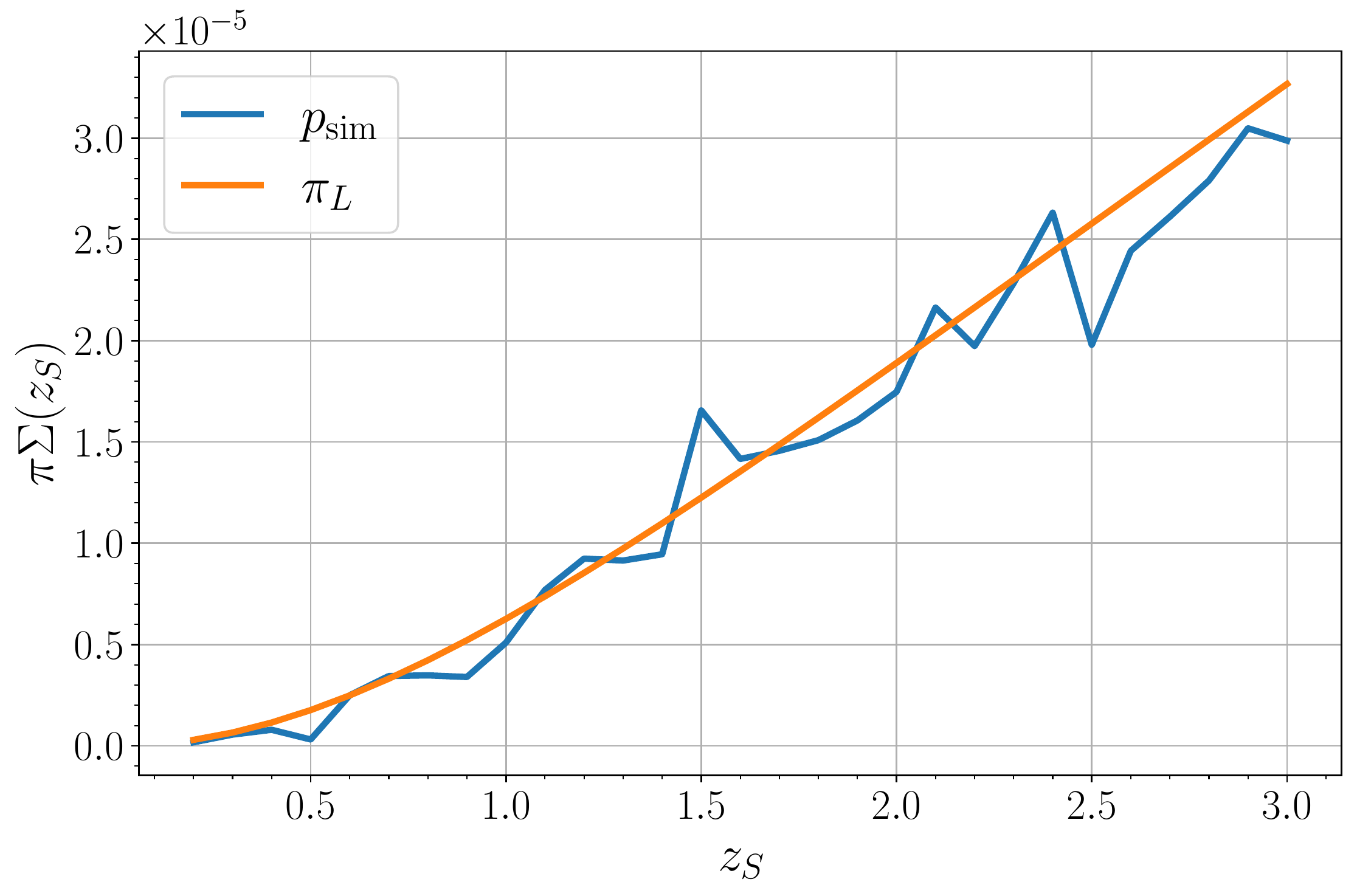}
    \caption{$\pi \Sigma(z_S, n_L = 1000 \; \perMpcCube)$ as a function of source redshift, comparing the analytical spatial Poisson model (orange) to simulated, fitted values (blue). As $\Sigma$ increases monotonically with source redshift, $\pi(y \vert z_S, n_L)$ shifts towards smaller $y$. Thus, detector networks with a larger detectable range are more likely to detect lensed sources.}
    \label{fig:alpha}
\end{figure}

\begin{figure}
    \centering
    \includegraphics[width=\columnwidth]{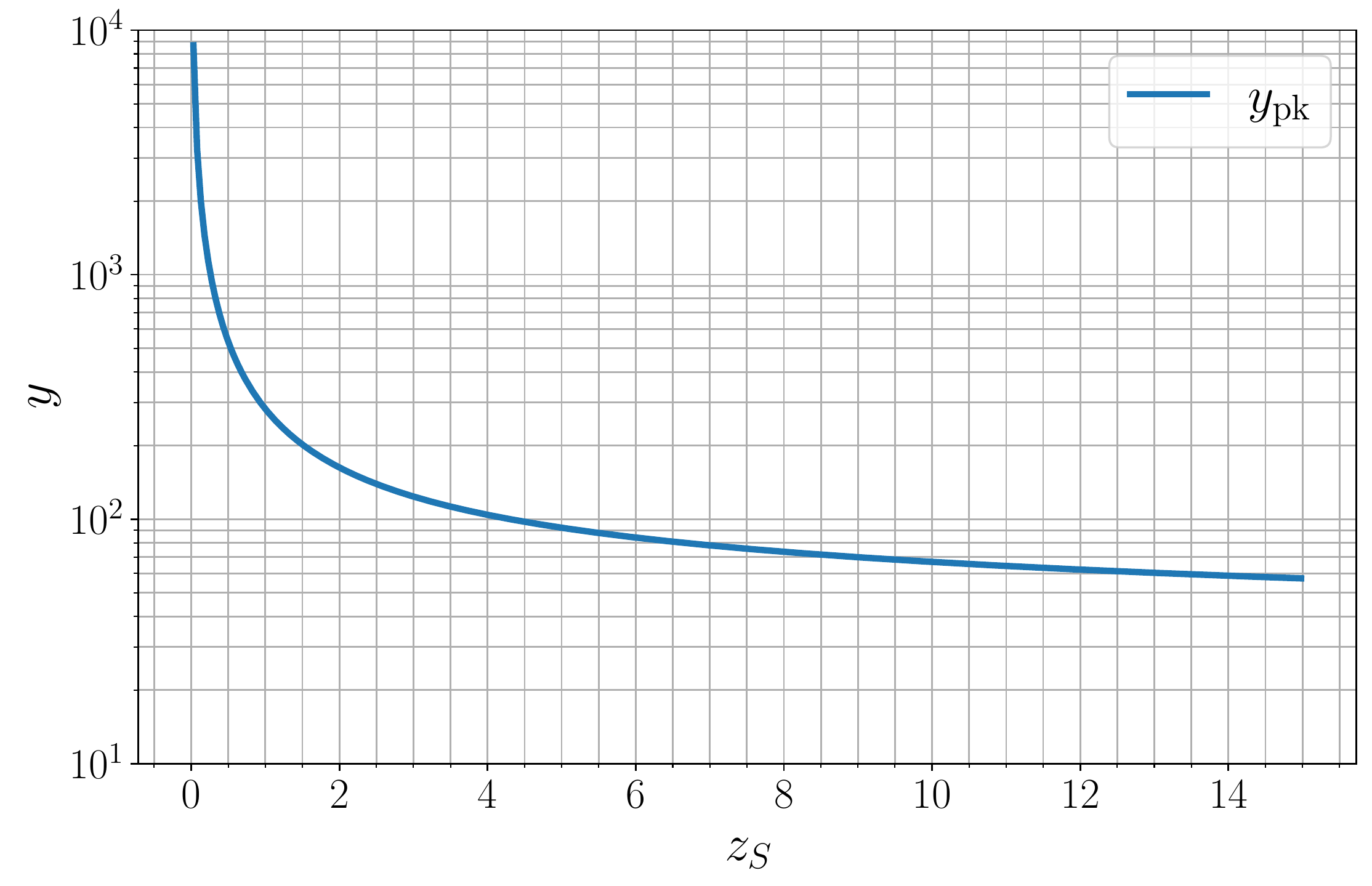}
    \caption{Peak $\pi(y \vert z_S)$ value as a function of $z_S$. The $y$ distribution shifts towards smaller values as the effective lens surface density grows.}
    \label{fig:y_pk}
\end{figure}

Fig. \ref{fig:alpha} shows the evolution of $\Sigma$ with source redshift. 
In particular, note that the effective density increases monotonically with source redshift, as more and more lenses are in the plane of the sky.
As a result, the $y$ distribution shifts towards smaller values as $z_S$ increases, and Fig. \ref{fig:y_pk} plots the decreasing peak value of $p(y|z_S, n_L)$ with $z_S$.

\section{Gravitational Wave Lens Parameter Estimation}
\label{sec:gwlens_pe}

\begin{figure}
    \centering
    \includegraphics[width=\columnwidth]{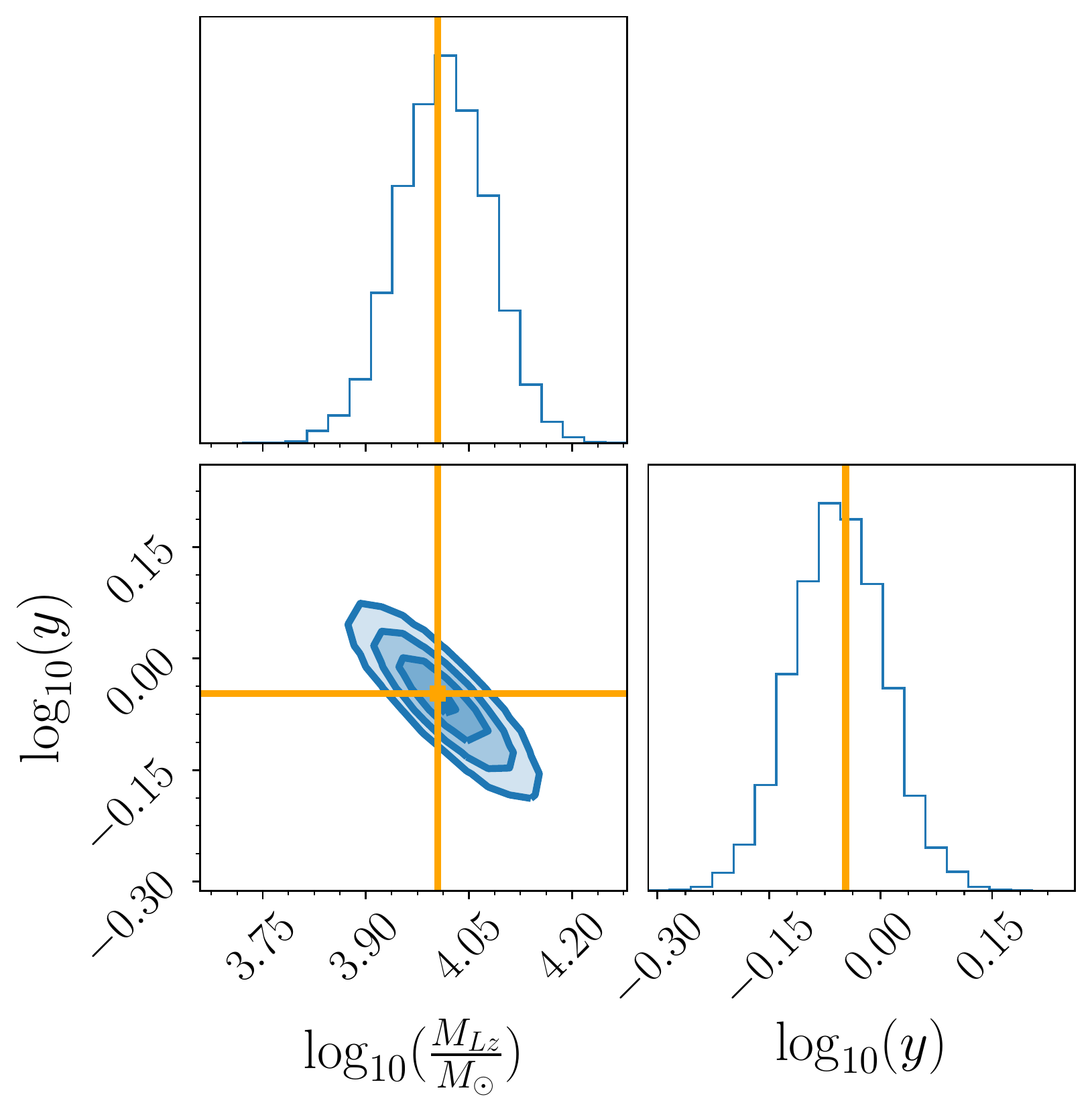}
    \caption{Corner plot posterior for an injection with $\mathrm{log}_{10}(y) \sim -0.05$ and $\mathrm{log}_{10}(M_{Lz}) \sim 4$, with the gold lines demarcating the injected values of $(M_{Lz}, y)$. While there is some degeneracy in $(M_{Lz}, y)$ as the parameter $w y^2$ determines the oscillatory behavior of the frequency domain waveform, both the injected $M_{Lz}$ and $y$ values are recovered with reasonable precision.}
    \label{fig:small_y_corner}
\end{figure}

\begin{figure}
    \centering
    \includegraphics[width=\columnwidth]{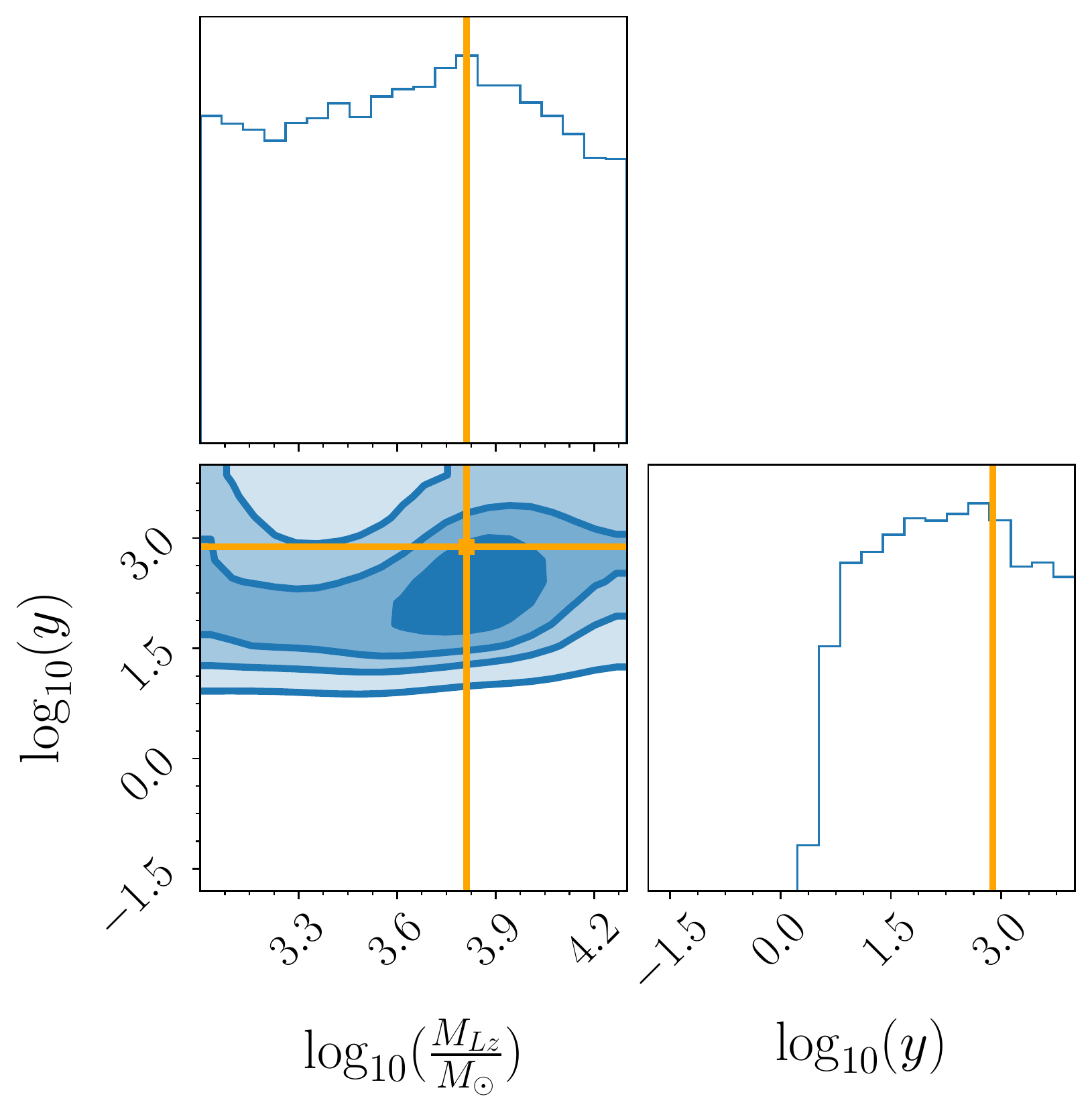}
    \caption{Corner plot with an injected $y \gg 1$ in $\{M_{Lz}, y \}$, with the injected values marked by the gold lines. In constrast to the small $y$ case, at large $y$ the lens mass posterior is completely agnostic, as is the $M_{Lz}$ posterior.}
    \label{fig:large_y_corner}
\end{figure}

In order to effectively use lens parameter estimation to draw conclusions on the IMBH population, injected lens parameters should be recoverable in the parameter estimation.
To conduct parameter estimation, we use the Bilby library \cite{ashton2019bilby} with the Dynesty sampler \cite{speagle2020dynesty}.
Fig.~\ref{fig:small_y_corner} and Fig.~\ref{fig:large_y_corner} demonstrate typical results for the impact parameter of a lensed gravitational wave injection, with an injected $y<1$ and $y \gg 1$ respectively.
In the case of $y<1$ in Fig. \ref{fig:small_y_corner}, the injected $y$ parameter is accurately recovered in the posterior of both $y$ and $M_{Lz}$, and the likelihood is only non-zero about the injected value.
Thus, injections with $y<1$ for IMBHs are clearly detectable.

In constrast to the small $y$ case, Fig. \ref{fig:large_y_corner} illustrates the posterior for a large injected value, $y \gg 1$.
With a uniform in log prior, the posterior remains relatively flat, and the posterior is \textit{not} localized about the injected value, as the effects of lensing on the waveform are too small to be detected, and the $M_{Lz}$ posterior is agnostic.
However, the posterior has no support for $y \lesssim 1$, ruling out the parameter space where lensing effects are significant.
In this way, the diffraction effects of a microlens can either be detected or ruled out.

At small $\nLo$ values the typical $y$ value is large, with the $y$ distribution peaking at $y_{p} \sim 1/\sqrt{\Sigma}$. 
This could present a problem if multiple diffraction effects are combined, as the lens with the smallest $y$ value for the source could be large enough that other lenses have a similar $y$ value.
However, as these parameter estimation results show, the diffraction effects are still minimal at large $y$, and so an arbitrarily large $y$ value can be injected without consideration of possible contaminating effects from other source-lens pairings in a multiple-lensing scenario.

\subsection{Generating the Injection Bank}

Finally, for a fixed lens number density and lens mass power law, we create an injection set to test our ability to recover the lens number density hyperparameter.
For the lens parameters, the source position $\pi(y \rvert n_L, z_S)$ is sampled from Eq.~\eqref{eq:pi_y}, and the source parameters are sampled from the distributions discussed in Sec.~\ref{app:source_pop}. 
For the base unlensed waveform, we use the IMRPhenomD approximate \cite{khan2016frequency, husa2016frequency}, which encompasses the inspiral, merger, and ringdown.
The lensed waveform is then the product of the amplification factor and the base waveform.
We threshold sampled injections by signal-to-noise ratio (SNR), selecting only those injections with network SNRs $\rho_{\rm{net}} > 12$ in a three detector network consisting of the LIGO Livingston, LIGO Hanford, and Virgo observatories at design sensitivity.

For the hyperparameters, we fix $\alpha_L = 1$, and generate injection sets with IMBH densities $n_L = \{10^3, 10^6\}  \mathrm{Mpc}^{-3}$.
At $n_L = \{10^3, 10^6\} \mathrm{Mpc}^{-3}$ the SNR gain due to strong lensing is negligible, and so we neglect the selection effect.

\section{Results of Hierarchical Analysis}
\label{sec:results}

\begin{figure}
    \centering
    \includegraphics[width=0.5 \textwidth]{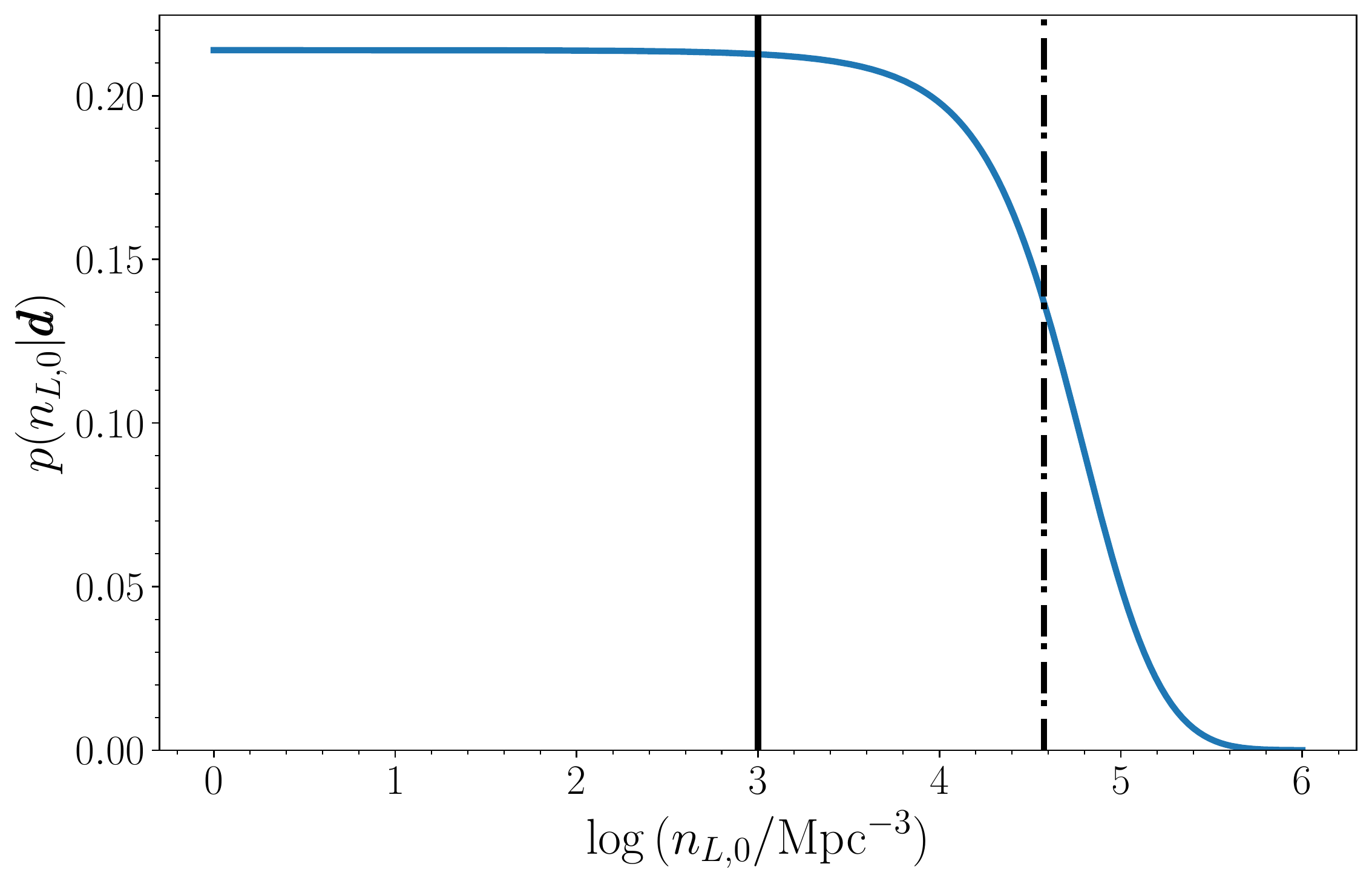}
    \caption{Hierarchical likelihood for an injected $n_L = 10^3~\perMpcCube$ density, with $\sim 200$ gravitational wave events. At $95 \%$ CI, the density is constrained to $\lesssim 10^{4.6}~\perMpcCube$. Further detections of gravitational waves unlensed by IMBHs could push this constraint further down, as well as an expanded, more sensitive detector network.}
    \label{fig:nl_1e3_results}
\end{figure}

\begin{figure}
    \centering
    \includegraphics[width=0.5 \textwidth]{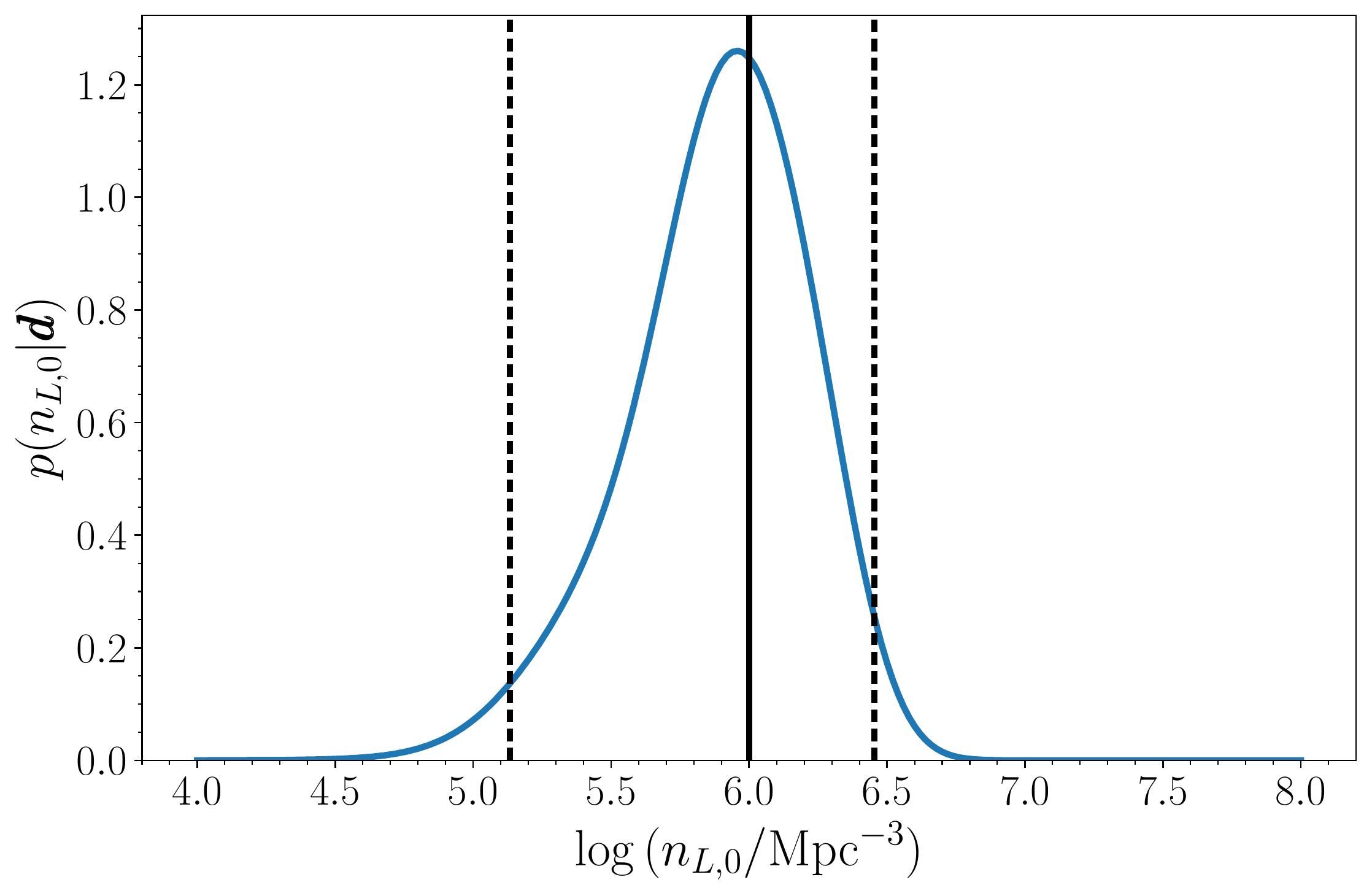}
    \caption{Hierarchical likelihood for an injected $n_L = 10^6 \mathrm{Mpc}^{-3}$ density, with $\sim 200$ gravitational wave events. The likelihood correctly recovers the injected hyperparameter $10^6 \perMpcCube$, with 95\% CI intervals of $10^{5.1}$ \perMpcCube and $10^{6.5}$ \perMpcCube, and so is capable of not only constraining the population properties of IMBHs but actually detecting them.}
    \label{fig:nl_1e6_results}
\end{figure}

Fig. \ref{fig:nl_1e3_results} and Fig. \ref{fig:nl_1e6_results} show the recovered hierarchical likelihood for the cases of $10^3 \mathrm{Mpc}^{-3}$ and $10^6 \mathrm{Mpc}^{-3}$ respectively. 
At $10^3 \mathrm{Mpc}^{-3}$, the recovered likelihood can constrain the hyperparameter to $\lesssim 10^5 \mathrm{Mpc}^{-3}$ at $90 \%$ confidence.
This upper constraint can improve with further unlensed detections.

For a density of $10^6 \mathrm{Mpc}^{-3}$, the injected hyperparameter is recoverable with this network, with the likelihood of Fig. \ref{fig:nl_1e6_results} ruling out both $n_L \lesssim 10^5 \mathrm{Mpc}^{-3}$ and $n_L \gtrsim 10^{6.5}  \mathrm{Mpc}^{-3}$ at $90 \%$ confidence.
Thus, even with just a three detector network, the population properties of IMBH lenses are not only possible to constrain but even to detect.
This is because $\mathcal{O}(1)$ events in our injection set are lensed with recoverable $y$ injection parameters in the parameter estimation, ruling out smaller lens number densities.

With a more sensitive network the volume of detectable mergers grows, and since $\yPriorFull$ increases monotonically with source redshift, the probability of encountering a significantly lensed event increases.
Thus, lensed events by IMBH lenses could be detectable even at these relatively small redshifts, and the recovered likelihood for an injected $10^3 \mathrm{Mpc}^{-3}$ hyperparameter may resemble a true measurement, rather than just an upper bound. 

\section{Discussion}
\label{sec:discussion}
We present a novel method of probing population distributions for lenses of gravitational waves, using the statistics of gravitational wave lensing, assuming that multiple lensing effects are negligible.
Deriving population models for the lensing statistics of point-mass lenses be distributed uniformly in comoving volume with a power-law mass distribution, we verify our models with direct simulations, and demonstrate a hierarchical Bayesian model for computing the likelihood of the lens density from successive observations.
We then conduct an injection campaign with gravitational wave samples, generating catalogues of lensed injections with network SNR $\rho_\mathrm{net} > 12$ for densities of $\{10^3, 10^6\} \mathrm{Mpc}^{-3}$.
Our results, shown in Figs. \ref{fig:nl_1e3_results}
and \ref{fig:nl_1e6_results}, show that we may either constrain or directly detect the lens number density for $\{10^3, 10^6\} \mathrm{Mpc}^{-3}$ respectively.

In the specific case of IMBHs, our method can probe their relatively unknown population properties with just a three-detector network of already existing gravitational wave observatories operating at design sensitivity.
Since the effective lensing probability increases with source redshift, a more sensitive detector network could greatly improve our ability to probe the IMBH population, detecting or constraining lower values of the lens number density.
With the addition of a few more planned observatories, like LIGO-India or KAGRA, the IMBH number densities of $\sim 10^3-10^4 \mathrm{Mpc}^{-3}$ could be directly detected.
Additionally, third generation detectors like the Einstein Telescope \cite{punturo2010einstein} or Cosmic Explorer \cite{abbott2017exploring, reitze2019cosmic} could probe extremely high source redshifts of $z_S \gtrsim 30$, detect $\sim 10000$ binary black hole mergers per month \cite{regimbau2017digging}, and be sensitive to higher injected $y$ values, so that smaller IMBH densities would be detectable.
Indeed, applying the third generation population forecast discussed in \cite{ng2020probing} with isolated galactic field formation, dynamical globular cluster formation, and Population III stars at high redshift subpopulations, we find that $\sim 1$ event with $y < 1$ could be detected each month for a density of $n_L = 10^3 \; \perMpcCube$.

We end by noting that the common use of lensing optical depth in Eq.~\eqref{eq:opticaldepth} carries the notion of a signal being \textit{lensed vs unlensed}, which is less well-defined in the wave-optics scenario.
The classification of the lensed signals relies on the choice of $y\leq y_{\rm max}$ to down-select the data of the lensed-only population for further analysis.
One has to build up detection statistics, e.g. the Bayes factor statistics from a large scale injection campaign~\cite{basak2021constraints} or the mismatch from the waveform~\cite{wang2021identifying}, for identifying the events that belong to the lensed population.
Besides being inflexible, this approach depends on a number of artificial choices, such as the choice of prior and the threshold of detection statistics for a lensed signal.
As a result, such process can be fuzzy for weak signals and may misidentify the lensed population in the data.
On the other hand, our method makes full use of the parameterization of $y$ and does not require the binary notion of ``lensed vs unlensed''.
With the hierarchical approach, we can treat the data as a whole population to infer the lens properties robustly, given a detailed model of the source-lens systems.

The mathematical framework derived in Secs.~\ref{sec:hba} \&~\ref{sec:lenspopulation} also allows for a flexible extension to test other lens models, such as the singular isothermal sphere or NFW profile~\cite{navarro1995simulations, navarro1996structure, navarro1997universal}, by considering the population as a mixture of different types of lenses.
Notably, inclusion of galactic lenses could boost the detectability of $y$ as shown in previous work \cite{seo2021strong}.
For lenses that do not obtain circular symmetry, such as elliptical lenses, the presented formalism still holds, with two modifications: (1) including the dependence of the symmetry-breaking parameter (e.g., ellipticity or external shear) in $\LensParam$ to calculate $F(f)$, and (2) redefining the normalization of $y$ that respects the notion of the nearest-effective lens, i.e., the effect of diffraction is stronger when $y$ is smaller, to evaluate $\Sigma$ and $\LensPriorFull$.
We will leave these extensions in the future work.

\section{Acknowledgements}
JG and ES are supported by grants from the Research Grants Council of the Hong Kong (Project No. CUHK 24304317), The Croucher Foundation of Hong Kong, and the Research Committee of the Chinese University of Hong Kong.
KKYN is supported by the NSF through the award PHY-1836814.
KWKW is supported by the
Simons Foundation.
The authors are grateful for computational resources provided by the LIGO Lab and supported by the National Science Foundation Grants No. PHY-0757058 and No. PHY-0823459. This research has made use of data, software and/or web tools obtained from the Gravitational Wave Open Science Center~\cite{vallisneri2015ligo}, a service of LIGO Laboratory, the LIGO Scientific Collaboration, and the Virgo Collaboration.

\appendix
\section{Amplification function in wave optics}
\label{appendix:a}
The background metric of a gravitational is given by
\begin{equation}
\mathrm{d}s^2=-(1+2U)\mathrm{d}t^2 +(1-2U)\mathrm{d}\textbf{r}^2 \equiv g^{(B)}_{\mu \nu} \mathrm{d}x^\mu \mathrm{d}x^\nu,
\end{equation}
with lens potential $U(\textbf{r}) \ll 1$. For a gravitational wave propagating against the lens background, we consider a linear perturbation against the background metric, where
\begin{equation}
    g_{\mu \nu} = g^{(B)}_{\mu \nu}+h_{\mu \nu}.
\end{equation}
Under an appropriate gauge choice and applying the Eikonal approximation, we can express the gravitational wave $h_{\mu \nu}$ as
\begin{equation}
    h_{\mu \nu} = \phi e_{\mu \nu},
\end{equation}
with polarization tensor $e_{\mu \nu}$ and scalar $\phi$. The change in the polarization tensor along the null geodesic is $\mathcal{O}(U) \ll 1$ such that we hold the polarization fixed. We then consider the propagation of the scalar field as it interacts with the background lens potential, with propagation equation
\begin{equation}
    \label{propeqn}
    \partial_\mu \left(\sqrt{-g^{(B)}}g^{(B)\mu\nu}\partial_\nu \phi\right)=0.
\end{equation}
In the frequency domain $\Tilde{\phi}(f, \textbf{r})$, Eq.~\eqref{propeqn} satisfies,
\begin{equation}
    \left(\nabla^2+\omega^2 \right)\Tilde{\phi} = 4 \omega^2 U \Tilde{\phi},
\end{equation}
where $\omega = 2 \pi f$.
We define the amplification function as the ratio of the lensed and unlensed ($U=0$) gravitational-wave amplitudes, such that 
\begin{equation}
    F(f)=\dfrac{\Tilde{\phi}^L(f)}{\Tilde{\phi}(f)}.
\end{equation}
In the thin-lens approximation, we decompose the source's wave into wavelets of all possible paths and integrate their contribution by the Kirchhoff's diffraction formula to obtain the amplification function~\cite{nakamura1999wave, nakamura1998gravitational, takahashi2003wave}
\begin{equation}
    \label{eq:diffracintegral}
    F(f)=\dfrac{D_S \xi_0^2(1+z_L)}{D_L D_{LS}}\dfrac{f}{i}\int \mathrm{d}^2 \pmb{x} \exp[2 \pi i f t_d(\pmb{x},\pmb{y})],
\end{equation}
where $D_S$ and $D_L$ are the source's and lens' angular diameter distances from the observer, respectively, $z_L$ is the lens redshift, $D_{LS}$ is the angular diameter distance between the source and lens, $\xi_0$ is the Einstein radius, $\bm{x}= \bm{\xi}/ \xi_0$ is the position of the wavelet on the lens plane, $\bm{y}=(\bm{\eta}/D_S)/(\xi_0/D_L)$ is the normalized impact parameter (or the normalized source position), and $t_d$ is the arrival time of the wavelet at the observer.
In the case of a point-mass lens, Eq. \eqref{eq:diffracintegral} may be analytically integrated yielding the solution 
\begin{align}
\label{eq:AFpointmass}
F(w)&=\exp\left\{\dfrac{\pi w}{4}+i \dfrac{w}{2} \left[\ln \left(\dfrac{w}{2}\right)-2\phi_m(y)\right]\right\} \nonumber \\
&\times \Gamma \left(1- \dfrac{i}{2}w\right) {_1}F_1\left(\dfrac{i}{2}w, 1; \dfrac{i}{2}w y^2\right),
\end{align}
where $w=8 \pi M_{Lz}f$ is the dimensionless frequency, $M_{Lz}$ is the redshifted lens mass, $_1F_1$ is the confluent hypergeometric function, and
\begin{align}
&\phi_m(y)=\frac{(x_m-y)^2}{2} - \ln x_m, \\
&x_m=\frac{y+\sqrt{y^2+4}}{2}.
\end{align}
To improve computational efficiency at the limit of $y \ll 1$ or $w\ll 1$, we switch to the geometric approximation of the magnification,
\begin{align}
    &F_{\rm geo}(w) = \sqrt{|\mu_{+}|} - i \sqrt{|\mu_{-}|} e^{i w \Delta \tau}, \\
    &\mu_{\pm} = \frac{1}{2} \pm \frac{y^2+2}{2y\sqrt{y^2+4}}, \\
    &\Delta \tau = \frac{y\sqrt{y^2+4}}{2} + \ln{\left(\frac{\sqrt{y^2+2}+y}{\sqrt{y^2+2}-y}\right)}
\end{align}
where $\mu_+$ and $\mu_-$ are the magnifications of the two geometric images, and $\Delta \tau$ is the normalized time delay between the two images.

\section{Source Distribution}\label{app:source_pop}
The parameters of the source distribution from which we sample are as follows.
For the mass distribution of the component source masses, we sample from the Power Law + Peak model from population studies of GWTC-2 \cite{abbott2020population}.
The source redshift distribution is drawn from the phenomenological fit to the population synthesis rate~\cite{Belczynski:2016obo,ng2020probing},
\begin{equation}
    \label{eq:sourceredshiftprior}
    p(z_S) \propto \frac{\di V_C}{\di z_S} \frac{(1+z_S)^{1.57}}{1+\left(\frac{1+z_S}{3.36}\right)^{5.83}}.
\end{equation}

The rest of the parameters, including the sky position, polarization angle, cosine of orbital inclination angle, and aligned spins, are distributed uniformly.
After sampling the source parameters from the above distribution, we simulate the gravitational-wave signals in the presence of detectors' noise, calculate the network SNR, and only select the signals with SNRs $\geq12$.

\bibliography{sample} 
\end{document}